# Evidence of robust 2D transport and Efros-Shklovskii variable range hopping in disordered topological insulator (Bi$_2$Se$_3$) nanowires

Biplab Bhattacharyya[1,2], Alka Sharma[1,2], Bhavesh Sinha[3], Kunjal Shah[3], Suhas Jejurikar[3], T. D. Senguttuvan[1,2] and Sudhir Husale[1, 2]*

[1] Academy of Scientific and Innovative Research (AcSIR), National Physical Laboratory, Council of Scientific and Industrial Research, Dr. K. S Krishnan Road, New Delhi-110012, India.

[2] National Physical Laboratory, Council of Scientific and Industrial Research, Dr. K. S Krishnan Road, New Delhi-110012, India.

[3] National Center for Nanosciences and Nanotechnology, University of Mumbai, Mumbai- 400098, India.

*E-mail: husalesc@nplindia.org

## Abstract

We report the experimental observation of variable range hopping conduction in focused-ion-beam (FIB) fabricated ultra-narrow nanowires of topological insulator (Bi$_2$Se$_3$). The value of the exponent $(d+1)^{-1}$ in the hopping equation was extracted as $\sim \frac{1}{2}$ for different widths of nanowires, which is the proof of the presence of Efros-Shklovskii hopping transport mechanism in a strongly disordered system. High localization lengths (0.5nm, 20nm) were calculated for the devices. A careful analysis of the temperature dependent fluctuations present in the magnetoresistance curves, using the standard Universal Conductance Fluctuation theory, indicates the presence of 2D topological surface states. Also, the surface state contribution to the conductance was found very close to one conductance quantum. We believe that our experimental findings shed light on the understanding of quantum transport in disordered topological insulator based nanostructures.

## Introduction

Quantum transport through topological insulator (TI) based nanostructures is among the most widely pursued research in recent condensed matter studies. The possibility of dissipationless electrical and spin current by topologically protected gapless surface states (SS) in TIs has encouraged researchers around the globe to shift towards one/two dimensional TI nanostructures, where due to large surface-to-volume ratio, contribution from topological surface states (TSS) is enhanced tremendously as compared to bulk.[1-9] However, complete isolation of bulk carriers is a very challenging task and several experimental works have been performed in the recent past to reduce the interference of bulk contribution in electrical transport studies in TI materials at low temperatures.[5,10-13] The nature of electronic transport highly depends on the synthesis/fabrication procedure of the TI-based nanodevice. Majority of the previous studies involved highly crystalline clean TI nanostructures like $Bi_2Se_3$ nanoribbons,[5] $Bi_2Te_3$ nanoribbons and nanowires,[6,14] $Sb_2Te_3$ nanowires,[10] SnTe nanowires,[15] etc to demonstrate the transport properties and quantum interference effects related to TSS, where the effect of disorder/localization on TSS properties was an unexplored territory. Some studies have highlighted the importance of electron-electron interaction (EEI) in quantum corrections to conductivity in addition to the weak antilocalization (WAL) in TIs.[16-19] Although, some theoretical models have focused on the effect of impurities, adatoms, strong electronic correlations and disorder on topological phase transition, electronic band structure, shifting of Dirac point due to induced surface potential, spin-orbit interaction and TSS conductance,[20-25] a very few experimental studies were performed on disordered and non-uniformly surfaced TI samples, which keep the electrical transport mechanism in such systems an interesting issue of discussion.

FIB fabricated TI-based nanowires provide an excellent way to study and understand the significance of disorder/localization on SS and bulk transport properties in TIs. Further, the constrained motion of non-trivial chiral fermions and contamination due to $Ga^+$ implantation, deformation of the material due to milling greatly affects the electronic correlations/interactions in the material. The theoretical works in the direction of impurity/disorder effects on TI properties have revealed that the TSS survives the weak to moderate perturbations (correlations/disorder/interactions), but under strong disorder, the topological properties show insulating characteristics.[23,26-28] In a recent study, it was demonstrated that introducing strong disorder in nanoscale TI samples might suppress the bulk conduction, while TSS remains immune to disorder.[29] Another report observed the change in band structure and TSS relocation to lower quintuple layers in presence of impurity and adsorbate addition.[20,21] The Universal conductance fluctuation (UCF) in weakly disordered TI specimens has also been a subject of intense research in recent years and different studies were performed to show the UCF arising from SS, the role of underlying symmetries (time reversal symmetry and spin rotational symmetry) in characterizing the UCF in 2D TIs, etc.[25,30-33] In another report, ion milling, which creates defect states (like vacancies) in semiconductors, was used to signify the role of hopping transport in the magnetoresistance (M-R) of TIs.[34] Hopping transport in disordered electronic systems is a manifestation of Anderson localization (AL).[35] Recently, Liao et al[36] reported the experimental demonstration of AL in ultrathin films of 3D TIs, where due to increasing the disorder, a crossover from WAL regime of diffusive transport to AL regime of Mott-type variable range hopping (VRH) transport was observed.

In this work, we present a comprehensive experimental analysis of VRH transport in highly disordered nanowires of $Bi_2Se_3$ of different widths, fabricated using the focused-ion-beam (FIB) technique. The excitation current and temperature dependent M-R curves of the $Bi_2Se_3$ nanowires were studied to know the robust nature of 2D transport in disordered TI nanowires. A very high resistance was observed for the nanodevices which showed Mott-type VRH mechanism. The VRH fitting of resistance versus temperature (R-T) curves, revealed a very less hopping potential barrier which is indicative of a high localization length. TSS contribution to conductivity, which is usually close to one conductance quantum ($e^2/h$), was observed dependent on the width of the nanowire. But the factor $p$ in the generalised VRH equation for ohmic regime,[37-39]

$$R(T) = R_o \, exp\left(\frac{T_o}{T}\right)^p \qquad (1)$$

is observed very close to $\frac{1}{2}$ for different widths of the nanowires indicating the no dependence on the dimensionality, i.e. the proof of the Efros-Shklovskii (ES-) VRH mechanism, which has been usually demonstrated in strongly disordered systems at low temperatures.[39-41]

**Results**

We have fabricated $Bi_2Se_3$ nanowire devices using the FIB milling technique from micro-mechanically exfoliated thin flakes of $Bi_2Se_3$ on cleaned $SiO_2$/Si substrates.[9,42] The details of the fabrication process are given in the Methods section. The standard four probe low temperature electrical transport measurements were carried out in Physical Property Measurement System (PPMS, Quantum Design) with a 9T magnet. In order to understand the effect of perpendicular magnetic field on the resistance of the FIB fabricated $Bi_2Se_3$ nanowire, first we have removed the contribution of UCF (discussed in the next section) by smoothing the M-R curve. For such a purpose, choosing a smoothing filter with small window size (~20 points)[43] and less order polynomial (~1 or 2) is better, as this would easily smooth out any extra unwanted oscillatory/fluctuating behaviour from the actual background signal. The orange coloured smoothed M-R curve can be seen along-with the actual M-R (in blue) at 2K (100nA current) for device W3 in Figure 1(a). A positive linear M-R with a WAL at zero magnetic field is observed, which is a characteristic of TI-based devices at low temperature.[5,14,44-46] In the past, a linear M-R was observed in nanoplates and nanostructures of $Bi_2Se_3$ and the M-R was found directly proportional to the carrier mobility (μ) and applied magnetic field (B), as

$$MR(T) = \mu(T)B \qquad (2)$$

The angular dependence measurements were performed on these $Bi_2Se_3$ nanostructures which confirmed the origin of linear MR as a signature of 2D transport property.[47-49] The linear M-R behaviour is also well explained with the classical model proposed by Parish and Littlewood.[50,51] Note that, a sharp WAL dip is not distinguishable in our nanowire, which can be attributed to the presence of other kinds of electronic interference/interaction effects originating from the disorder/defect sites in our samples, which compete with the quantum interference effect causing WAL. The background curve (orange) in Figure 1(a) shows the oscillations at higher magnetic fields. Earlier reports have assigned such oscillation in TI materials under perpendicular field to the Shubnikov de-Haas (SdH) effect.[6,11,14] In order to understand the origin of SdH oscillations in our TI-based sample, we performed the Landau level fan diagram analysis. It is a well-known fact that with increase in the magnetic field, successive emptying of Landau levels (LL) takes place, which manifests itself as oscillations in the resistance for a 2D Fermi surface.[11] The LL index ($n$) is related to the magnetic field ($B$) and Fermi cross-sectional area ($S_F$) by

$$2\pi(n + \gamma) = S_F \frac{\hbar}{eB} \qquad (3)$$

where $S_F = \pi k_F^2$, $e$ = electronic charge and $\hbar$ = reduced Planck's constant. The parameter of interest in the above equation is the factor $\gamma$, which is known as the Berry's phase and its value is predicted to be ±0.5 for a TSS. The value of $\gamma$ can be calculated from the Y-intercept in LL index versus $B^{-1}$ plot as shown in Figure 1(b), where the red circles and olive squares represent the corresponding minima and maxima in M-R oscillations (inset in Fig. 1(b)), respectively. The value of Y-intercept in our case comes out to be -0.54, which is very close to the $1/2$-shifted SdH oscillation predicted for TSS. This analysis indicates towards a very important property of TSS, i.e. robustness to disorder/deformation. The presence of large UCF superimposed on the background M-R did not affect the $1/2$-shifted energy spectrum related to the TSS. Also, both the Landau level fan diagram and UCF (as discussed in next section) analysis suggests the presence of TSS in our samples.

Fluctuations in the resistance as a function of magnetic field for device W3 (length ~ 1.846 μm, width ~ 110.3 nm) are clearly evident from Figure 2. We performed the temperature and excitation current (at a particular temperature) dependent M-R analysis, in order to understand the origin of these fluctuations. Figure 2(a) shows the change in M-R, i.e. $\frac{R(B)-R(0T)}{R(0T)} \times 100\%$, curves at 100nA excitation current for temperatures 2, 4 and 6K. The curves have been shifted for better clarity. The decreasing amplitude of these fluctuations with temperature, reproducibility in reverse magnetic field sweeps and the aperiodicity with respect to magnetic field suggested us to compare this effect with the UCF. It is a well known phenomenon that the electronic interference in disordered mesoscopic metallic samples with sizes smaller than the phase coherent length $(L_\varphi)$ manifests itself as some noisy/aperiodic patterns of the order of $e^2/h$ in the M-R curves, known as UCF.[52,53] When the sample size becomes larger than $L_\varphi$, self-averaging effects come into play, which decreases the amplitude of fluctuations.[53,54] Inset of Figure 2(a) shows the fluctuations in conductance, which has been extracted by subtracting the background conductance, obtained from a fit to a five order polynomial, from actual conductance. Although, complete reproducibility of fluctuations at different temperatures is not expected in our case due to very high disorder and deformations in the nanowire due to milling procedure, a good amount of reproducible features are present in the fluctuation curves of 2K and 4K (shifted for clarity in inset of Figure 2(a)), which reflect the inherent fingerprint of the sample. Quantitative analysis of the amplitude of fluctuations in our sample was performed using the UCF theory proposed by Lee et al.[53] The conductance fluctuations for UCF analysis can be obtained from the relation:

$$\delta G = G(B) - \langle G(B) \rangle \qquad (4)$$

where $\langle G(B) \rangle$ denotes the ensemble average of magneto-conductance. A correlation function of the fluctuations

$$\mathcal{F}(\Delta B) = \langle \delta G(B) . \delta G(B + \Delta B) \rangle \qquad (5)$$

is used to calculate the amplitude of UCF, i.e., root mean square value of the conductance fluctuation ($rms(\delta G)$ or $\sqrt{\langle \delta G^2 \rangle}$), at $\sqrt{\mathcal{F}(0)}$.[53] The calculated UCF magnitude at different temperatures is around 0.01 $e^2/h$ and depicts a slow exponentially decaying dependence with temperature ($T^{-0.47}$) as shown in the lower inset of Figure 2(b). Here, an important quantity of interest is the full width at half maximum of the auto-correlation function

$$\mathcal{F}(B_c) = \langle \delta G(B) . \delta G(B + B_c) \rangle \qquad (6)$$

which is known as the correlation field $B_c$ and can be calculated using the relation $\mathcal{F}(B_c) \cong \frac{1}{2}\mathcal{F}(0)$.[32,33,53] The upper inset in Figure 2(b) shows the extracted $B_c$ values as a function of the temperature for our case (See Methods section for $B_c$ calculations). A power law relationship $B_c \sim T^{0.71}$ is obtained by fitting with an exponential function. Further, it is known that in a 2D system, when $L_\varphi$ is the shortest length, we have

$$B_c(T) \sim \frac{\phi_o}{L_\varphi^2} \qquad (7)$$

where $\phi_o$ is equal to the magnetic flux quantum, $h/e$.[53] Using equation (7), we calculated the $L_\varphi$ values and its temperature dependence (Fig. 2(b)) show an exponential decay with $L_\varphi \sim T^{-0.36}$. Theoretically, the power law dependence of $L_\varphi$ for a 1D, 2D and 3D system is predicted to be proportional to $T^{-1/3}$, $T^{-1/2}$ and $T^{-3/4}$, respectively.[55] For a 2D system with $L_\varphi < L_z$ and $L_\varphi < L_x$ we have

$$rms(\delta G) \sim \left[\frac{L_x}{L_z}\right]^{1/2} \left[\frac{L_\varphi}{L_z}\right] = \left[\frac{L_x^{1/2}}{L_z^{3/2}}\right] \times L_\varphi \qquad (8)$$

where $L_\varphi$ is the phase coherent length and $L_x$, $L_z$ are width and channel length of the sample.[53] Since $L_\varphi \sim T^{-0.5}$ for a 2D system, therefore from Eq. (8), $rms(\delta G) \sim L_\varphi \sim T^{-0.5}$. The temperature dependence of UCF amplitude in our case is $T^{-0.47}$, which is very similar to the theoretically expected characteristics[31,33,53,54] and suggests a 2D dominated system. Note that we did not observe the theoretically accurate estimate predicated for 2D or 1D transport, i.e. UCF amplitude $T^{-0.5}$ or $T^{-0.33}$, respectively. Earlier studies demonstrating the 2D transport in TI materials have also reported small deviations in comparison to the theoretical values. Kim et al.[32] obtained a relation of $L_\varphi \sim T^{-0.43}$ for Bismuth nanowires, which reflected that the system is a hybridization of both one and two dimensions. Since, the exponent of -0.43 was closer to the 2D system value of -0.5, therefore, it was considered that the SS plays a major role in transport. Trivedi et al.[33] reported slightly deviating values of T$^{-0.44}$ and T$^{-0.6}$ from $rms(\delta G)$ and $L_\varphi$ analysis in the nanosheets of Bismuth Telluro-Sulfide, and compared the case with 2D system. Matsuo et al.[30] reported a dependence of T$^{-0.65}$ for quasi-one-dimensional $Bi_2Se_3$ nanowires, where again the transport was attributed to the 2D system. We believe that further investigations on ultra-narrow wires (widths < 100nm) are required to shed more light on the nature 1D transport.

The excitation current dependent UCF study at 2K has been shown in Figure 3(a) and (b). The decrease in M-R and conductance fluctuations with increasing current is clearly evident from Figure 3(a), which shows the M-R change curves at different currents of 100nA, 500nA and 1.8μA. The average conductance almost doubles itself from $0.7\ e^2/h$ to $1.3\ e^2/h$, while increasing the current from 5nA to 1.8μA, shown in Figure 3(b) (right Y-axis). This is indicative of the fact that the transmission probability of carriers between the electrodes and the nanowire is dependent on the current bias. A possible explanation for this could be the carrier confinement effects introduced due to the low dimensionality of the fabricated nanowire. Figure 3(b) (left Y-axis) shows the exponential decrease in $rms(\delta G)$ with current, i.e. $rms(\delta G) \sim I^{-0.44}$. Since, the voltage can be considered proportional to the current in our case, therefore, the relationship becomes $rms(\delta G) \sim V^{-0.44}$. Clearly, in our sample, the amplitude of conductance fluctuation drops with voltage as $1/\sqrt{V}$. The dependence of UCF on voltage (or current) bias in mesoscopic conductors is a highly complicated phenomenon and some studies in the past were conducted to understand it. Ludwig et al[56] found that for non-interacting electronic systems, the variance $\langle \delta G^2 \rangle$ shows a monotonic increase with voltage, i.e., an enhancement in fluctuations in observed with increase in voltage, which is in good agreement with the pioneering work by Larkin and Khmel'nitskii.[57] Simultaneously, they also discussed about the systems with high EEI, where dephasing due to Coulomb interaction occurs and results into a non-monotonic dependence of conductance fluctuations with voltage. Further, it was shown that in such systems $rms(\delta G)$ falls as $1/\sqrt{V}$ for higher voltages. In our case, the FIB fabricated nanowire resembles to a strongly disordered electronic system with sufficiently high localized states, leading to high EEI and dephasing effects, which may be a reason for the observed $1/\sqrt{V}$ drop in the amplitude of conductance fluctuations.

## Discussion

We now discuss the non-trivial effects of disorder and localization induced hopping transport in the $Bi_2Se_3$ nanowires fabricated by using FIB milling method. For this purpose, we performed the detailed analysis of the temperature dependence of conductance for device W3 and W6. The upper right inset in Figure 4(a) shows the cooling curve for device W3. A huge monotonic increase in the resistance with decreasing temperature till 4.4K is observed. Below this, the resistance increases very sharply from 18kΩ (at 4.4K) to 19.6kΩ (at 2K). The R-T behaviour in our sample clearly indicates a huge insulating behaviour. Previous TI-based transport studies have reported metallic or semiconducting R-T behaviour based on the bulk conduction band or shallow impurity band.[5,6,11,58] In our case, none of these standard models properly fit to the experimental results. Therefore, we use the VRH model to fit our obtained results. Figure 4(a) shows the VRH fit (blue solid line) on conductance versus temperature plot (orange circles) for device W3. In addition to the standard Mott-type VRH conductance, a constant conductance is also used to correctly define the overall conductance of the sample.[29] Therefore, the total conductance ($G$) is given by the relation:

$$G(T) = G_{TSS} + G_o \exp\left[-\left(\frac{T_o}{T}\right)^{1/d+1}\right] \quad (9)$$

where $G_{TSS}$ is the conductance contribution from TSS, $G_o$ is a pre-factor, $T_o$ is a characteristic temperature related to the hopping energy and $d$ is the effective dimension of the system. The exponent $(d+1)^{-1}$ determines the type of conduction mechanism in the sample based on the shape of density of states (DOS) at the Fermi level.[40,41] A recent study on $Bi_2Te_3$ nanotubes showed that the temperature dependent second term in Eq. (9) comes from the bulk channel.[29] A closer inspection of Figure 4(a) suggests that the experimental conductance deviates from the VRH fit at ~ 42K (shown in the blue shaded region). The deviation from VRH transport mechanism and the sharp decrease in conductance can be attributed to the high EEI in TIs at low temperatures, which is discussed after the hopping mechanism. The blue solid line is a fit to Eq. (9) using fitting parameters $G_{TSS} \sim 6 \times 10^{-5}$ S, $G_o \sim 1.68 \times 10^{-4}$ S, $d \sim 1.3$ and $T_o \sim 940.7$ K for device W3. Figure 4(b) demonstrates that the TSS contribution is a constant slightly above one conductance quantum ($e^2/h \sim 3.874 \times 10^{-5}$ S) for most of the temperature range but decreases for temperatures below 6K due to EEI effects. The TSS conductance was obtained by subtracting the bulk conductance contribution given by the second term in Eq. (9) from the experimentally acquired total conductance. Previous works have shown similar kind of TSS contribution close to one conductance quantum.[29,59] Also, a comparison between the total conductance and bulk conductance is plotted in inset of Figure 4(b), which clearly reflects that the bulk contribution is very low compared to the total conductance, which is evident from the low value of $G_o$. A similar kind of analysis was performed on another nanowire device W6 having much greater width (~ 683 nm) than W3. FESEM image of the device is shown in the upper right inset of Figure 4(c). The lower inset shows the R-T characteristic of the device. Again, we observe a highly insulating behaviour for this device with resistance increasing up-to 33kΩ at 2K. A very nice fit of the VRH equation (red solid line in Figure 4(c) and upper left inset) for the whole temperature range is obtained for this device. The values of fitting parameters obtained for W6 are $G_{TSS} \sim 1.1 \times 10^{-5}$ S, $G_o \sim 2.44 \times 10^{-4}$ S, $d \sim 1.5$ and $T_o \sim 23.42$ K. Figure 4(d) indicates that the contribution from bulk conductance is very large and almost equal to the total conductance. Thus, the TSS contribution is very less than one conductance quantum and is shown by the fluctuating $G_{TSS}$ values of ~ $0.3\, e^2/h$ in inset of Figure 4(d). The above two results clearly indicate that the conduction through TSS is dominant in W3 nanowire, where the surface-to-volume ratio is very high as compared to W6. This result is very interesting as it demonstrates the FIB fabrication method as an efficient way to tune the conductance contribution from SS and bulk transport. We observed that there are two kinds of conductance channels, one from the TSS and another from the bulk. With increasing width of the nanowire the contribution from bulk also increases due to more conduction channels available in the bulk. Previously, Ref. [29] stated that due to strong disorder in the system bulk conductance is suppressed and TSS contribution is enhanced significantly for TI nanotubes with very less cross-sectional area. A similar kind of phenomenon is observed in our case for smaller width nanowire W3.

The deviation from VRH model was observed in the conductance at very low temperature for device W3. Here we use the EEI theory to explain this phenomenon. The nanowire conductance falls off logarithmically with temperature below 5K (lower inset in Figure 4(a)). The correction in 2D conductance due to EEI is given by the relation: [16,17,60]

$$\Delta G_{EEI}(T) = \frac{e^2}{2\pi^2 \hbar}\left(1 - \frac{3}{4}\tilde{F}_\sigma\right) ln\left(\frac{T}{T_o}\right) \qquad (10)$$

where $\tilde{F}_\sigma$ is the electron screening factor and $T_o$ is the reference temperature from where the correction $\Delta\sigma$ is measured. The values obtained for $\tilde{F}_\sigma$ and $T_o$ from the fitting curve (solid purple line in lower inset of Figure 4(a)) are 0.2 and 2.1K, respectively. Theoretically, it is predicted that the value of $\tilde{F}_\sigma$ lies in the range of 0 to 1, and a value close to zero implies the presence of strong EEI in the system. Many previous reports have exhibited the relevance of many-body interactions in low-dimensional TI samples. The anomalous insulating behaviour in TIs at low temperatures was proposed as a result of co-existence of both the strong electronic interactions and the topological delocalization (TD).[16,19] The enhancement of Coulombic interaction could suppress the density of states at the Fermi level and thus, a logarithmic increase in resistance is observed for device W3. Also, the competition between the TD and EEI at low temperature modifies the quantum correction to conductance as

$$\Delta G(T) = \Delta G_{TD}(T) + \Delta G_{EEI}(T) \qquad (11)$$

where $\Delta G_{TD}(T)$ is the enhancement in conductance by TD and $\Delta G_{EEI}(T)$ is the suppression of conductance by EEI.[16,18] The short phase coherence length and less WAL observed in our nanowire can be attributed to the strong electronic interaction, which is enhanced in thinner samples due to disorder and reduced dimensionality. Previous reports have shown that in very thin TI samples an energy gap opens up at the Dirac point due to the coupling of top and bottom surfaces, which weakens the TD and reduces conductance at low temperature.[61,62]

We believe that our experimental findings are a consequence of the strong localization effects created in the sample by disorder/deformation, and FIB milling procedure is inherently well known to introduce such defects/deformations/disorder. In presence of disorder, localized/defect states are created and electron transport occurs by hopping across these states which is a totally different behaviour than that of other conduction electrons.[63] The DOS around the Fermi level is an important parameter for understanding the hopping of electrons between the localized states. In our case, the VRH analysis for both the nanowires gives almost the same value of exponent $(d + 1)^{-1}$, which is close to $\frac{1}{2}$. There is no dependency of hopping transport on the width/dimensions of the nanowire. This is indicative of the fact that under high disorder regime, ES-VRH mechanism is followed in our nanowires, where the DOS near the Fermi level is not constant (in contrast with Mott-VRH where DOS is constant near the Fermi level) and vanishes linearly with energy, and the value of the exponent is $\frac{1}{2}$ in all dimensions.[39,40] In a previous work on $Bi_2Se_3$ thin films, the characteristic temperature $T_o$ in Eq. (9) was found to be ~ $10^6$ K.[64] But in our case, it is 940.7 K and 23.42 K for W3 and W6, respectively. Since, the hopping potential is proportional to $T_o$, we can conclude that the energy required by electron to hop from one localized state to another is very small in our case. Also, the probability of hopping,

$$P \sim exp\left(-2\alpha R - \frac{W}{kT}\right) \qquad (12)$$

where $R$ is the spatial separation and $W$ is the energy separation between two localized states,[65] is more for our case. The localization length was calculated using the expression

$$\xi = \frac{2.8\, e^2}{4\pi\varepsilon\varepsilon_o k_B T_o} \qquad (13)$$

where $\varepsilon_o$ and $\varepsilon$ represent the permittivity of vacuum and the dielectric constant of the material.[39,41] The value of $\varepsilon\varepsilon_o$ is taken as $100\varepsilon_o$ for $Bi_2Se_3$.[66,67] High localization lengths of 0.5nm and 20nm for W3 and W6, respectively, indicate the presence of strong disorder in our nanowires. Due to high localization and confinement effects in our nanowire, $L_\varphi$ tends to be very small in comparison to the dimensions of the nanowire. Also, our analysis suggests that for a low dimensional disordered TI sample, voltage induced dephasing takes place, so that $L_\varphi$ is unstable and varies with the excitation current. It is very important to note that the conductance fluctuations in our case cannot be simply attributed to the noise due to current induced heating effects, since our nanodevice was perfectly placed in liquid Helium with a proper heat exchange setup. Also, the heating effects due to ramping up of magnetic field can be ruled out, as ramping rates of less than 30 Oe/sec were used. The UCFs have a totally different origin from other quantum interference effects like WAL. The amplitude of UCF provides an excellent way to study the quantum transport properties in a material, since UCF is related to the phase-coherent length, sample size and symmetry of the Hamiltonian inside the sample.[68] Recently, quantum oscillations in the form of UCF were not only used to reveal the topological character of the material but also to study the phase transition from Dirac-to-Weyl semimetal by breaking time reversal symmetry in $Cd_3As_2$ nanowires.[69] In the case of 3D TIs, both the bulk and SS contribute to the amplitude of UCF.[31] The absence of backscattering in TIs suppresses the number of scattering paths and thus, leads to the scattering confinement, which plays a significant role in 2D UCF.[70] Previously, for TI samples a $T^{-1/2}$ dependence of $L_\varphi$ were considered as an important signature of 2D TSS.[45] However, in our case the electronic transport seems to be originating from the hybridization of 1D and 2D system, as indicated by the exponent of -0.36. The unusual exponent close to 1D behaviour is of no surprise, since the fabricated nanowire in our case resembles a quasi-1D system. It is important to note that the quantum confinement effects due to low dimensionality of the system also affect the transport properties in a TI-based system, and many future studies need to be performed in order to fully understand the complex mechanism related to the interplay of 2D TSS and quantum confinement effects.

**In summary**, we have experimentally demonstrated the existence of ES-VRH mechanism in highly disordered/deformed nanowires of $Bi_2Se_3$. The device fabrication procedure used in our case inherently provides an excellent way to test for the robustness of TSS to strong disorder introduced in the system via $Ga^+$ implantation, deformation formed due to milling and other impurities deposited during in situ metal contacts formation. Our analysis revealed that the UCFs originate from the 2D-TSS, thus confirming the robust nature of TSS. Also, it was shown that the contribution of TSS and bulk can be studied in the FIB fabricated nanowires of $Bi_2Se_3$.

## Methods

**Device fabrication.** The thin $Bi_2Se_3$ nanowire based devices were fabricated by micromechanical cleavage technique using the standard scotch tape method and focused ion beam (FIB) milling. Firstly, $SiO_2$/ Si wafers (p-type highly Boron doped with ~0.001-0.005 $\Omega$-cm) were cleaned with acetone, iso-propanol, methanol and de-ionised water, and additional oxygen plasma treatment was performed for ~10min. The thin flakes of $Bi_2Se_3$ (99.999% CAS#12068-69-8, Alfa Aesar) were deposited using exfoliation method $SiO_2$/ Si substrates with pre-sputtered thick Au/Ti pads (~80/5 nm). This method produces thin and random sized flakes of $Bi_2Se_3$ which were further localized under optical microscope (Olympus MX51) and field emission scanning electron microscopy (FESEM, Zeiss-Auriga). The located thin flakes were milled using FIB, with $Ga^+$ ions as a source material for the ion beam. The electrical contacts on the thin nanowire of $Bi_2Se_3$ were made with metal electrodes of platinum through FIB based gas injection system (GIS).

**Correlation field ($B_c$) calculation.** The correlation function $\mathcal{F}(B_c) = \langle \delta G(B) \cdot \delta G(B + B_c) \rangle$ was used to extract the $B_c$ values at different temperatures. $\delta G$ can be calculated using the relation $\delta G = G(B) - \langle G(B) \rangle$. Firstly, the Taylor series was used to expand the function $\delta G(B + B_c)$ as:

$$\delta G(B + B_c) = \delta G(B) + \frac{B_c^1}{1!}\delta G'(B) + \frac{B_c^2}{2!}\delta G''(B) + \frac{B_c^3}{3!}\delta G^{(3)}(B) + \frac{B_c^4}{4!}\delta G^{(4)}(B) + O(B_c^5)$$

For better accuracy, we include the first five terms (till $O(B_c^4)$) on the right hand side of the above equation. The derivatives of $\delta G$ were calculated and used as an input for the above equation along-with a $B_c$ value. The mean of the Hadamard product of $\delta G(B)$ and $\delta G(B + B_c)$ gives the correlation function $\mathcal{F}(B_c)$. In order to verify the correctness of $B_c$ value, we use the relation $\mathcal{F}(B_c) \cong \frac{1}{2}\mathcal{F}(0)$, where $\mathcal{F}(0) = \langle \delta G^2 \rangle = rms^2(\delta G)$.

## Acknowledgments

B.B. acknowledges the fellowship of Council of Scientific and Industrial Research, India. S.H. and A.S. acknowledge CSIR's Network project "Aquarius" for the financial support.

## Author contributions

A.S. performed sputtering and exfoliation process. S.H. designed, supervised the research and fabricated the nanodevices. B.S., B.B., S.J. and K.S. conducted the transport measurements. T.D.S. provided FIB instrument operation support, tools and materials. B.B. and S.H. analysed the data and wrote the manuscript.

## Additional information

Reprints and permissions information is available online at www.nature.com/reprints. Correspondence and requests for materials should be addressed to S.H.

## Competing financial interests

The authors declare no competing financial interests.

## Corresponding Author

*E-mail: husalesc@nplindia.org

# **Figures**

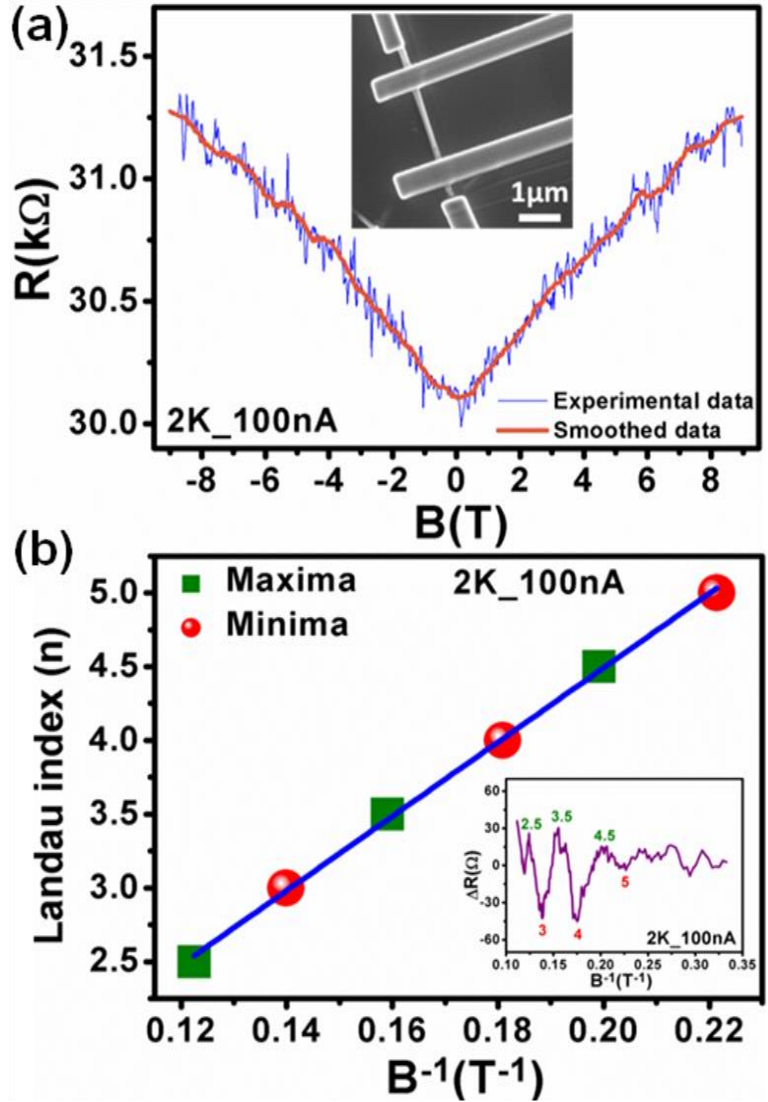

**Figure 1. Landau level fan diagram analysis.** (a) M-R for device W3 shows a positive fluctuating behaviour at 2K (in blue). The smoothed M-R (shown in red colour) depicts oscillatory behaviour in the background at high magnetic fields. Inset shows the FESEM image of the nanowire device. (b) Landau fan diagram for the smoothed M-R curve, with maxima and minima positions as shown in the inset.

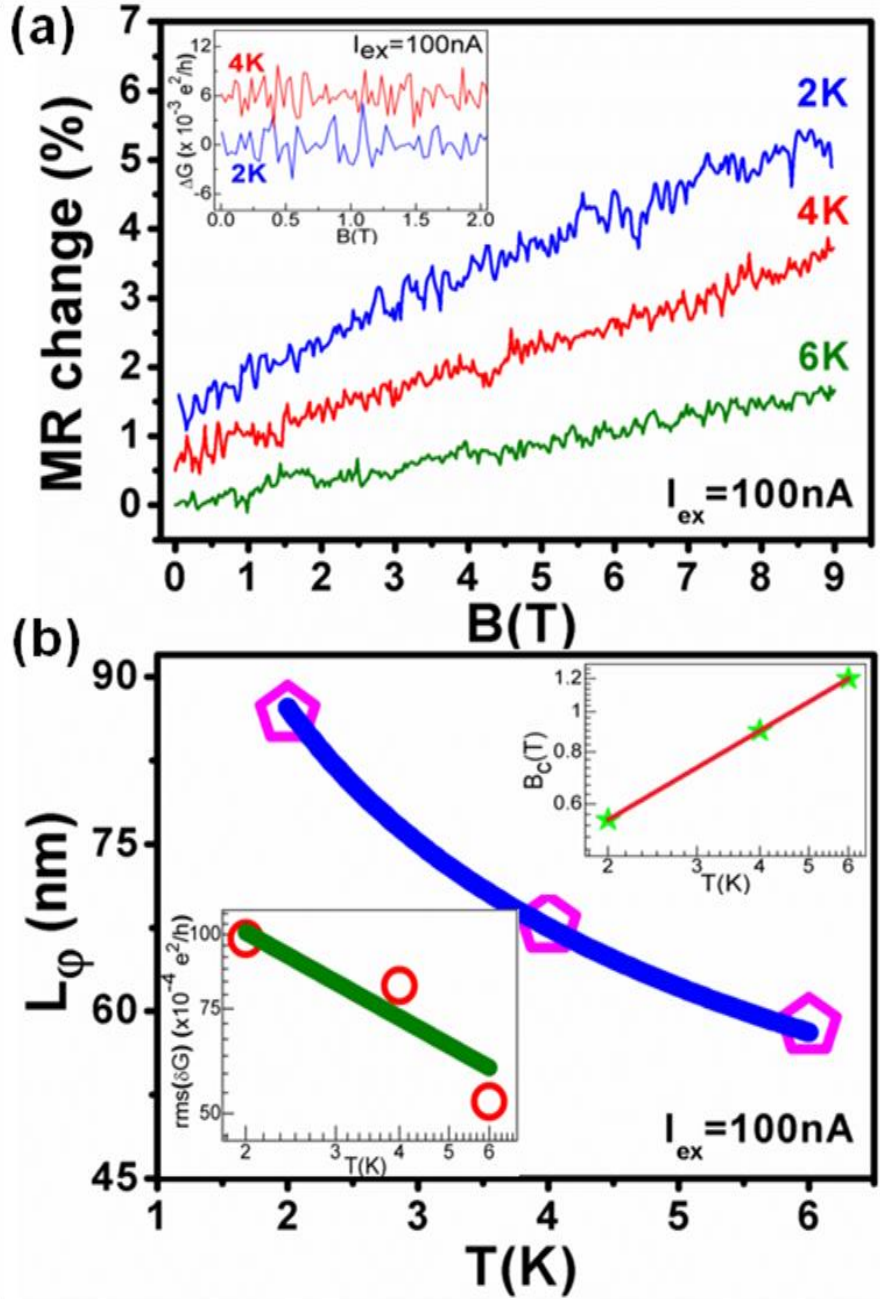

**Figure 2. Temperature dependence of UCF for device W3.** (a) M-R change (%) curves at different temperatures. Resistance shows clear fluctuations with magnetic field for different temperatures at 100nA excitation current. Amplitude of fluctuations gets reduced with increase in temperature. Inset shows the reproducible M-R fluctuations for different temperatures. (b) $T^{-0.36}$ phase coherence length decay as a function of temperature. Lower inset shows the slowly decaying UCF magnitude with temperature. An exponential function of $rms(\delta G) = cT^a$ was used, where $c = 0.014 \pm 0.003$ and $a = -0.47$ was extracted from the fit. Upper inset shows the correlation field values calculated using correlation function at different temperatures.

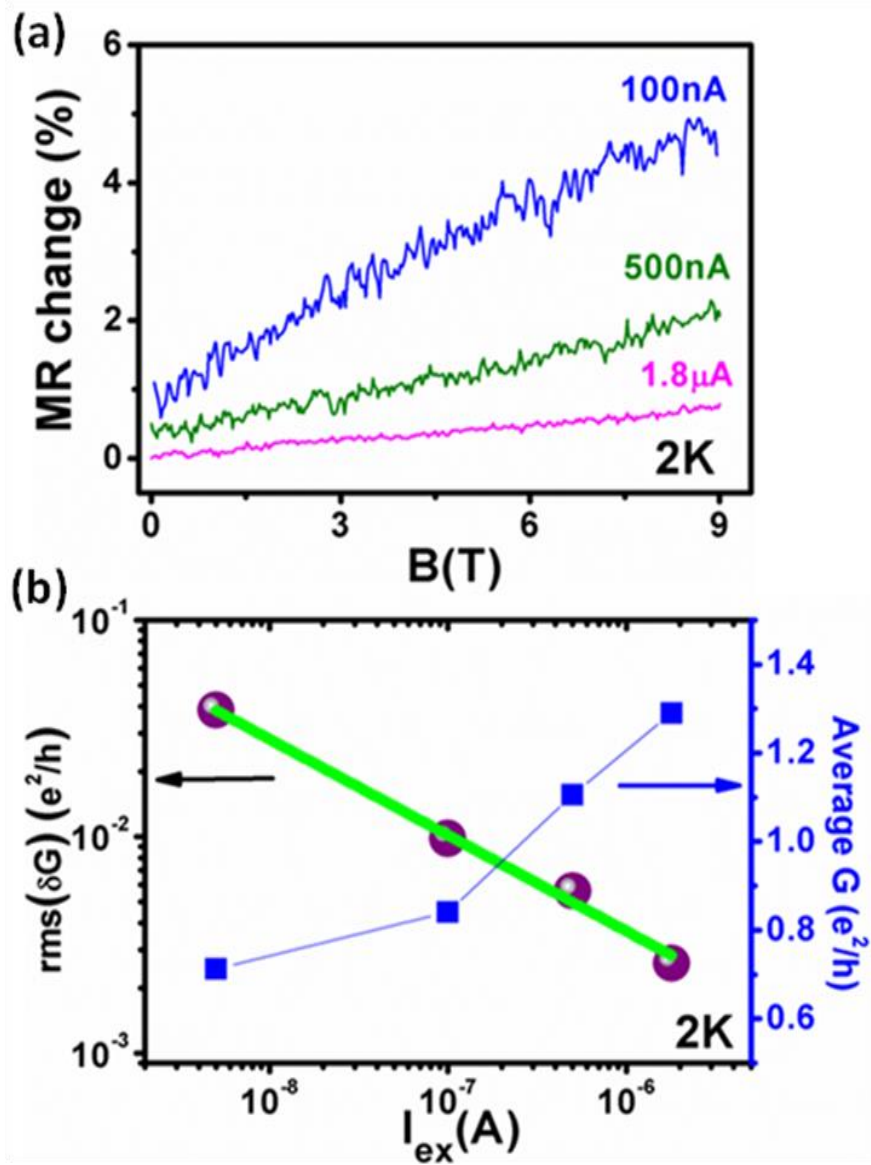

**Figure 3. Current dependence of UCF for device W3.** (a) M-R change (%) curves for different excitation currents at 2K. A decrease in M-R fluctuations is observed with increasing excitation current. (b) Left panel shows the exponentially decreasing fluctuation amplitude with excitation current. Right panel shows the increase in average conductance with excitation current.

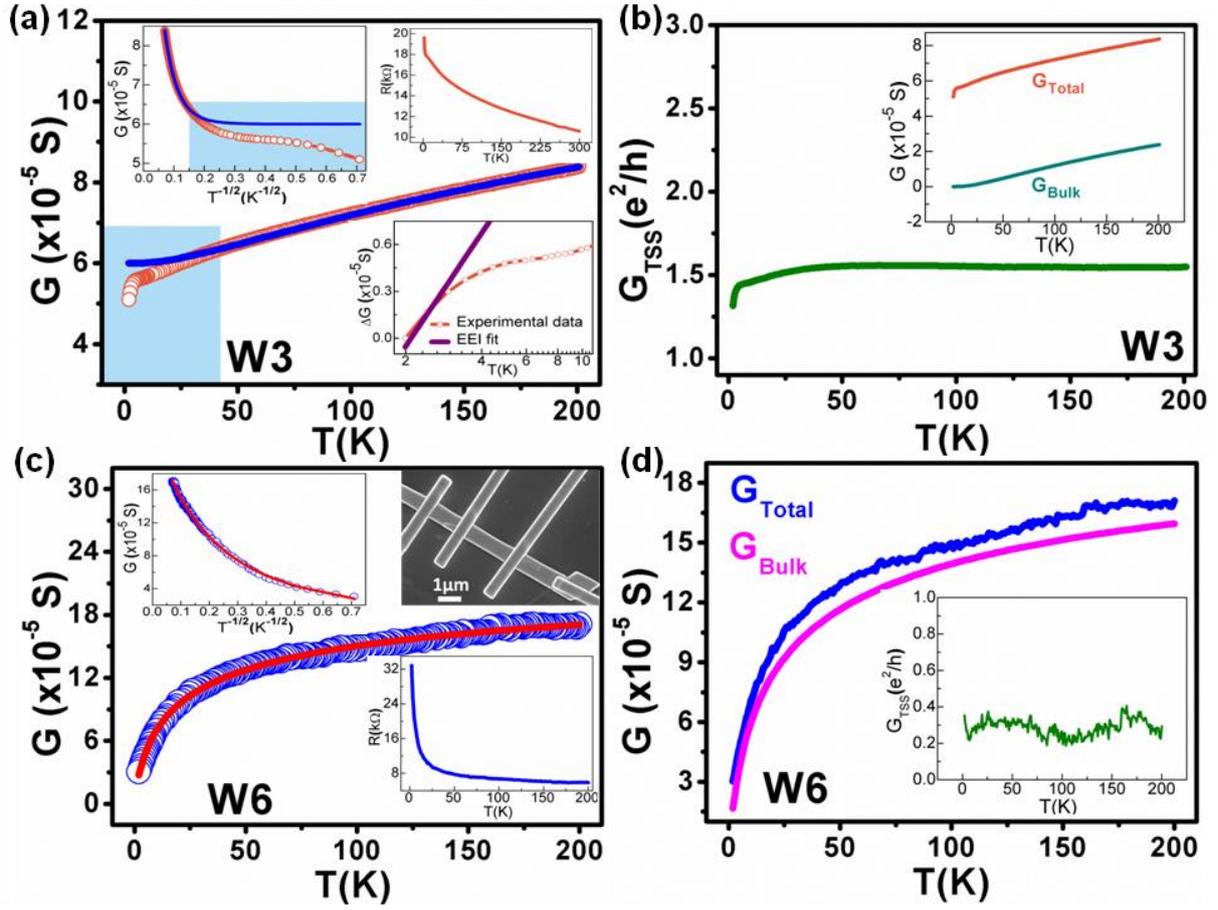

**Figure 4. Variable range hopping transport mechanism in $Bi_2Se_3$ nanowires under strong disorder regime.** (a) Conductance versus temperature plot for device W3 (orange circles) with VRH fitting (blue solid line). Upper right inset shows the R-T cooling curve and lower inset shows the EEI fit at very low temperatures. Upper left inset is the G vs. $T^{-1/2}$ plot. The blue shaded region depicts the deviation in experimental conductance from VRH model. (b) TSS contribution to conductance is just above one conductance quantum ($e^2/h$) for device W3. Inset shows the comparison between total conductance and bulk conductance. (c) Conductance (blue circles) for device W6 shows a monotonic decrease for complete temperature range. Upper left inset shows the G vs. $T^{-1/2}$ behaviour. VRH fitting is shown in red colour. Lower inset shows the R-T cooling curve for the device. Upper shows the FESEM image of W6. (d) High bulk contribution is observed for the device W6. The TSS conductance is very low (~ 0.3 $e^2/h$).